\PassOptionsToPackage{bookmarks=true}{hyperref}
\documentclass[sigconf]{acmart}

\usepackage{booktabs}
\usepackage{url}
\usepackage{caption} 
\usepackage{xspace}
\usepackage{paralist}
\usepackage{subfig} 
\newcommand{\mRUBiS}{{\mbox{mRUBiS}}\xspace}
\newcommand{\CompArch}{{\mbox{CompArch}}\xspace}
\newcommand{\cf}{{\textit{cf.}}\xspace}
\newcommand{\eg}{{\textit{e.g.}}\xspace}
\newcommand{\ie}{{\textit{i.e.}}\xspace}
\newcommand{\elem}[1]{\textsf{#1}}
\newcommand{\artifact}[1]{\textit{\textbf{{#1.}\xspace}}}
\newcommand{\projectURL}{\url{https://github.com/thomas-vogel/mRUBiS}}

\copyrightyear{2018}
\acmYear{2018}
\setcopyright{rightsretained}
\acmConference[SEAMS '18]{SEAMS '18: 13th International Symposium on Software Engineering for Adaptive and Self-Managing Systems}{May 28--29, 2018}{Gothenburg, Sweden}
\acmBooktitle{SEAMS '18: 13th International Symposium on Software Engineering for Adaptive and Self-Managing Systems, May 28--29, 2018, Gothenburg, Sweden}
\acmDOI{10.1145/3194133.3194161}
\acmISBN{978-1-4503-5715-9/18/05}
\fancypagestyle{arxiv}{\fancyhf{}\fancyfoot[C]{~\\~\\This is the author's version of the work. It is posted here for your personal use. Not for redistribution. The Definitive Version of Record was published in SEAMS'18, May 28--29, 2018, Gothenburg, Sweden. DOI: \href{https://doi.org/10.1145/3194133.3194161}{10.1145/3194133.3194161}.}}

\begin{document}
\title{mRUBiS: An Exemplar for Model-Based Architectural Self-Healing and Self-Optimization}

\renewcommand{\shorttitle}{mRUBiS: An Exemplar for Model-Based Architectural Self-Adaptation}

\author{Thomas Vogel}
\orcid{0000-0002-7127-352X}
\affiliation{%
	\institution{Humboldt-Universit\"{a}t zu Berlin, Germany}
}
\email{thomas.vogel@cs.hu-berlin.de}


\begin{abstract}
Self-adaptive software systems are often structured into an adaptation engine that manages an adaptable software by operating on a runtime model that represents the architecture of the software (model-based architectural self-adaptation). Despite the popularity of such approaches, existing exemplars provide application programming interfaces but no runtime model to develop adaptation engines. Consequently, there does not exist any exemplar that supports developing, evaluating, and comparing model-based self-adaptation off the shelf. Therefore, we present \mRUBiS, an extensible exemplar for model-based architectural self-healing and self-optimization. \mRUBiS simulates the adaptable software and therefore provides and maintains an architectural runtime model of the software, which can be directly used by adaptation engines to realize and perform self-adaptation. Particularly, \mRUBiS supports injecting issues into the model, which should be handled by self-adaptation, and validating the model to assess the self-adaptation. Finally, \mRUBiS allows developers to explore variants of adaptation engines (e.g., event-driven self-adaptation) and to evaluate the effectiveness, efficiency, and scalability of the engines.
\end{abstract}

%
%

\ccsdesc[500]{Software and its engineering~Development frameworks and environments}
\ccsdesc[500]{Software and its engineering~Software development techniques}

\keywords{Self-adaptation, architecture, runtime models, simulator}

\maketitle
\thispagestyle{arxiv}
\section{Introduction}

Self-adaptive software is a closed-loop system that implements a feedback loop to autonomously adapt to changes at runtime~\cite{2013-SEFSAS2a}. Engineering such systems is challenging as it involves domain and adaptation concerns.
Thus, a popular engineering approach is to split the system in two parts: an \textit{adaptation engine} that implements the feedback loop and an \textit{adaptable software} that realizes the domain logic while the engine manages the software~\cite{Salehie&Tahvildari2009,Weyns+2012}. This split promotes separation of concerns but requires that the engine has a representation of the software based on which it can perform adaptation.
Such a representation is a \textit{runtime model} that is causally connected to the adaptable software, that is, changes of the software are synchronized to the model and vice versa~\cite{MC.2009.326}.
According to the survey by \citet{Salehie&Tahvildari2009}, most self-adaptive systems in research follow such a \textit{model-based self-adaptation} approach.
In this context, many researchers argue that the \textit{software architecture} is the appropriate abstraction level for self-adaptation and thus for the runtime model~\cite{Magee+Kramer1996,
Oreizy+1998,
Garlan+Schmerl2002,
Dashofy+2002,
GarCHSS04,
Bradbury+2004,
Floch+2006,
Kramer&Magee2007,
Morin+2009,
Edwards+2009,
2009-ICAC,
2010-SEAMS,
Weyns+2012b,
Angelopoulos+2013,
Braberman+2017,
Mahdavi-Hezavehi+2017}.  
Consequently, our focus is on \textit{model-based architectural self-adaptation}.

To advance software engineering research, benchmarks and exemplars have been identified as useful~\cite{Sim+2003}. Researchers have proposed such artifacts to develop and evaluate self-adaptation solutions, which can be categorized in two groups.
\begin{inparaenum}
\item Artifacts that support developing and evaluating self-adaptive software and that range from
requirements for the adaptive internet of things~\cite{FeedMe},
component frameworks for smart cyber-physical systems~\cite{DEECo,IntelligentEnsembles},
a platform for cloud applications~\cite{Hogna},
a tool for runtime monitoring and verification~\cite{Lotus},
and 
a benchmark environment for self-adaptive applications in Hadoop clusters~\cite{HadoopBenchmark}.
\item Several exemplars that provide a real or simulated adaptable software, on top of which adaptation engines should be developed, and that enable evaluation and comparison of adaptation engines~\cite{Znn,ATRP,TAS,SAVE,UNDERSEA,DeltaIoT,CrowdNav}.
\end{inparaenum}

Despite the popularity of model-based architectural self-adap\-tation, none of these artifacts particularly addresses this kind of self-adaptation by providing an architectural runtime model of the adaptable software.
Artifacts of the first group typically do not provide a runtime model. Exceptions are Hogna~\cite{Hogna} and Lotus~\cite{Lotus} that use a performance model for performance analysis respectively a labeled transition system model for verification. However, these models cannot be used for generic, architectural self-adaptation.
The exemplars of the second group provide application programming interfaces (APIs) to the adaptable software instead of a runtime model.
Adaptation engines developed for these exemplars use these APIs to manage the software. Consequently, using these exemplars for model-based architectural self-adaptation requires from developers to implement a runtime model and a causal connection between the model and the APIs. 
This is challenging since developers have to assure the synchronization and fidelity of the runtime model with the running software~\cite{2014-MART,MC.2009.326}.
Therefore, we conclude that there does not exist any exemplar that supports developing, evaluating, and comparing model-based architectural self-adaptation off the~shelf.

In this paper, we present \textit{\mRUBiS}\footnote{The exemplar is available at \projectURL.}, an extensible exemplar for model-based architectural self-adaptation. It simulates the \mRUBiS system as the adaptable software and provides as well as maintains an \textit{architectural runtime model} of the system. This model serves as the interface for adaptation engines to realize and perform architectural adaptation of \mRUBiS. 
Thus, model-based architectural self-adaptation is supported off the shelf. Developers are relieved from implementing a runtime model and a causal connection to the adaptable software as well as setting up a corresponding runtime infrastructure. Instead, they can focus on designing, implementing, and evaluating the adaptation logic on top of the provided runtime model. With respect to developing self-adaptive software, such an approach supports the early testing of feedback loops~\cite{2015-SASOW}.
 
The simulation performed by the exemplar consists of a predefined number of iterations over the following three steps:
\begin{inparaenum}[(1)]
\item According to a \textit{scenario}, the simulator injects \textit{issues} into the runtime model and thus to the \mRUBiS architecture, which should be handled by self-adaptation.
\item The adaptation engine developed by the user of the exemplar is triggered to analyze and adapt the \mRUBiS architecture described in the model. The adaptation aims at resolving the injected issues and thus at satisfying the goals of \mRUBiS. 
\item According to a set of \textit{validators}, the simulator validates the adaptation and runtime model to check whether issues are remaining in the \mRUBiS architecture. 
It further evaluates the self-adaptation by computing the utility of the current architecture based on a \textit{utility function} and by measuring the execution time of the self-adaptation (\ie, of the 2nd step). This data is summarized at the simulation end.
\end{inparaenum} 

For the \mRUBiS architecture, we provide scenarios, issues, validators, and utility functions for a self-healing and a self-optimization case study. Furthermore, each of these elements can be replaced or extended by developers to address other case studies. Even \mRUBiS as the adaptable software described by the runtime model can be replaced or extended. The \textit{\CompArch} language to express the runtime model is generic and supports modeling arbitrary component-based architectures and properties. This guarantees the extensibility of the exemplar although prescribing the language for the model.

However, the exemplar does not restrict the adaptation engines developed on top of it. In contrast, it even encourages developers to explore variants of engines, for instance, by using the provided change events to drive the self-adaptation. Developers can use their favorite technologies to implement the engines such as code (Java) or model-based rules (\eg, expressed with OCL and Story Diagrams) that operate on the runtime model.
Finally, the exemplar allows developers to evaluate the effectiveness (in terms of the utility of the adaptable software), efficiency (in terms of execution time), and scalability of self-adaptation by scaling the size of the architectural model and the number of injected issues per simulation round. 

The paper is structured as follows: 
Sec.\,\ref{sec:background} presents the background. We discuss the generic, \mRUBiS-specific, and implementation aspects of the exemplar in Sec.\,\ref{sec:generic-framework},~\ref{sec:mRUBiS}, and~\ref{sec:implementation} before concluding in Sec.\,\ref{sec:conclusion}.

\section{Background: Models at Runtime}
\label{sec:background}

\begin{figure}[htb]
	\centering
	\includegraphics[scale=.39]{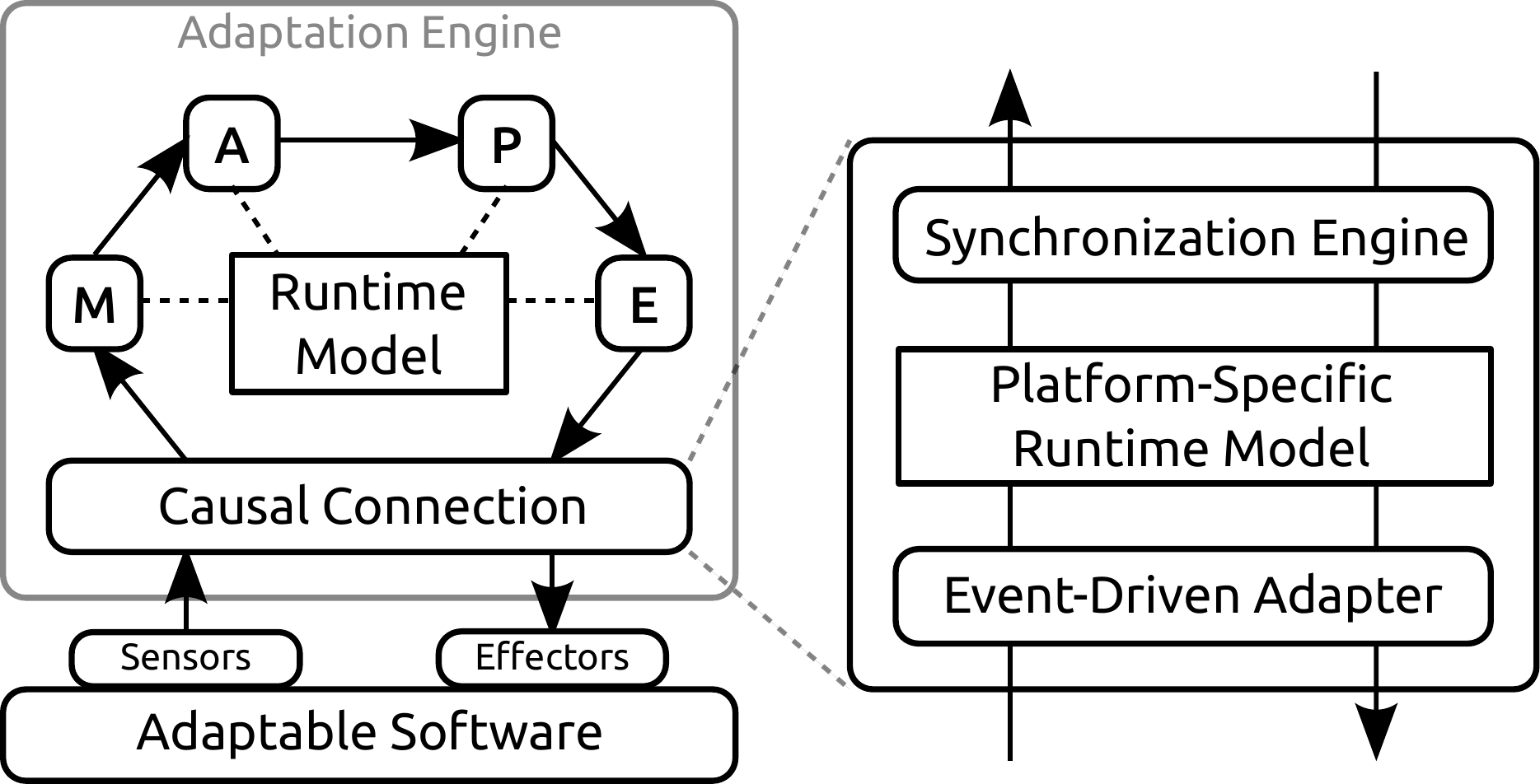}
	\caption{Self-adaptive system with a runtime model.}
	\label{fig:mart}
	\vspace{-1.5em}
\end{figure}

In this section, we conceptually outline the use of a runtime model by a self-adaptive system (\cf~Figure~\ref{fig:mart}). 
Although the \mRUBiS exemplar does not impose a certain structure on the adaptation engine, we use the MAPE structure~\cite{Kephart&Chess2003} for the illustration.

As shown in Figure~\ref{fig:mart}, the self-adaptive system is split into an adaptation engine and adaptable software. The engine realizes the four MAPE steps that operate on the runtime model to perform adaptation. Such an approach requires a \textit{causal connection} that propagates changes of the adaptable software to the runtime model and changes of the model prescribing an adaptation to the software (\cf~\cite{MC.2009.326,2014-MART}). Thus, the monitoring and execution steps use the causal connection to synchronize the model and the adaptable software.

However, implementing a causal connection is challenging as developers have to address the abstraction gap between the software and the architectural runtime model, assure the fidelity of the model with the running software, and achieve a runtime-efficient synchronization~\cite{2014-MART,MC.2009.326}. 
These aspects make the implementation of a causal connection a costly task considering, for instance, the complexity of the Rainbow solution that uses an architectural model of the Znn exemplar for self-adaptation~\cite{GarCHSS04,Znn}.
With a similar complexity, we realized the causal connection in earlier work~\cite{2009-ICAC,2010-SEAMS} (\cf Figure~\ref{fig:mart}). Using the sensor and effector APIs, an event-driven adapter incrementally maintains a platform-specific runtime model that is at the same abstraction level as the APIs. A synchronization engine then propagates changes between the platform-specific and architectural runtime model taking the abstraction gap into~account.

Based on these observations, we may conclude that developing self-adaptive systems with runtime models is challenging as it causes accidental complexity of implementing a causal connection. In contrast, developers should rather focus on the essence, which is developing the adaptation logic and particularly the analysis and planning mechanisms.
Therefore, with the \mRUBiS exemplar we propose simulating the adaptable software and causal connection so that developers experimenting with adaptation engines are relieved from implementing a runtime model and causal connection.

\section{Generic Simulation Framework}
\label{sec:generic-framework}

We now present the generic simulation framework that underlies the exemplar and that is independent of any adaptable software.

\subsection{Overview}
\label{subsec:overview}

\begin{figure}[htb]
	\vspace{-1em}
	\centering
	\subfloat[Structure.\label{fig:simulator}]{\includegraphics[scale=.39]{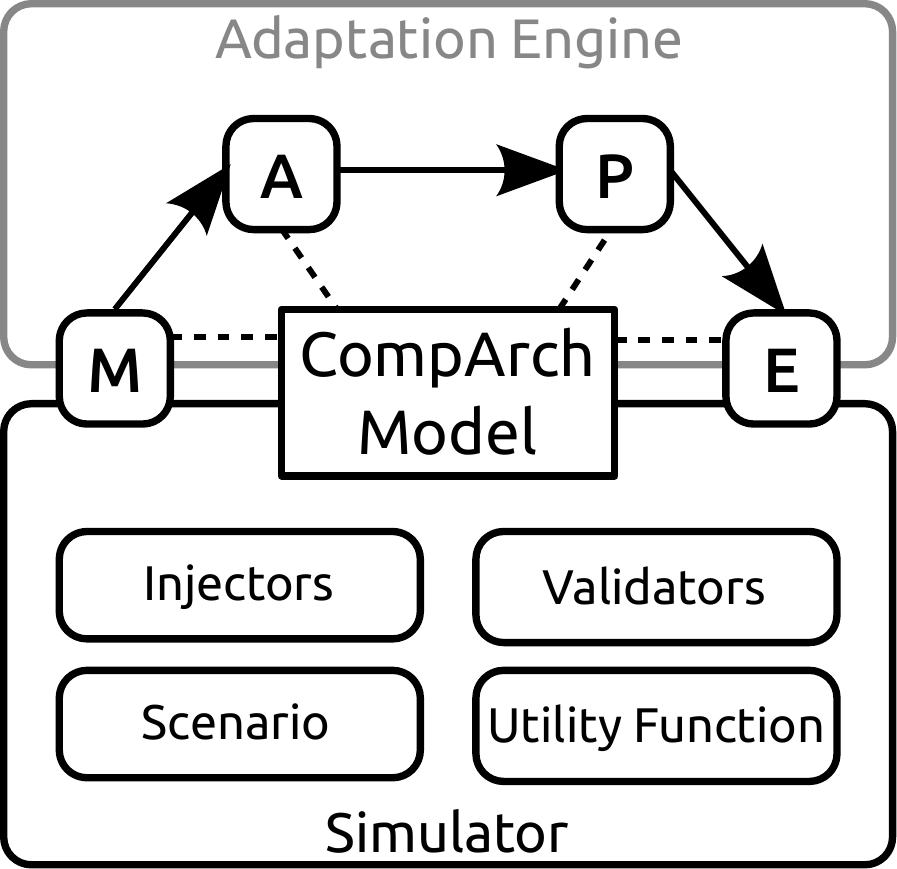}} 
	\hspace{0.02\linewidth}
	\subfloat[Simulation steps.\label{fig:simulation}]{\includegraphics[scale=.43]{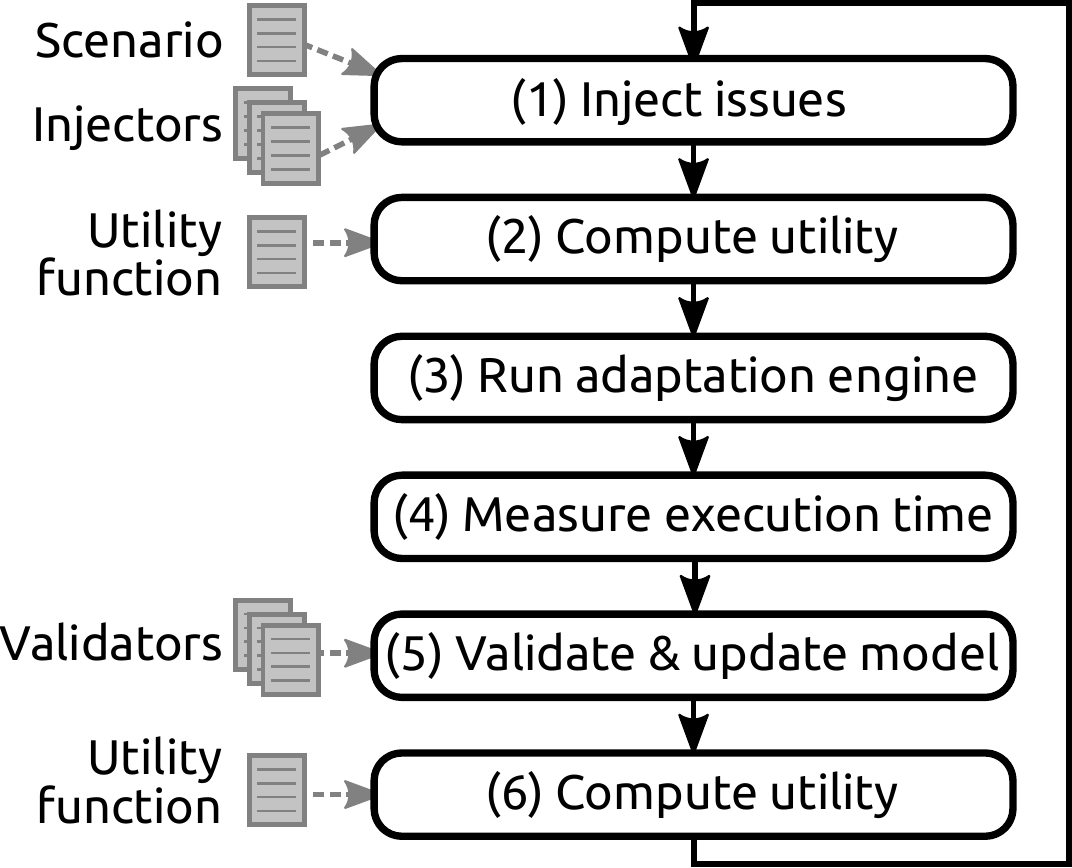}} 
	\caption{Generic simulation framework.}
	\label{fig:framework}
	\vspace{-1em}
\end{figure}

The structure of the generic simulation framework is shown in Figure~\ref{fig:simulator}. As previously motivated, the simulator emulates the adaptable software and causal connection by providing an architectural runtime model, the \textit{\CompArch Model} (\cf Figures~\ref{fig:mart} and~\ref{fig:simulator}). This model is expressed in the generic \textit{\CompArch} language (\cf Section~\ref{subsec:comparch}). Thus, the simulator takes over the part of the monitoring and execution steps of a feedback loop that would otherwise realize the causal connection.
The adaptation engine developed by users of the exemplar only relies on the \CompArch model without having to use any sensor or effector APIs. Thus, the MAPE steps directly use and operate on the \CompArch model to perform self-adaptation.

To enable the simulation, the simulator must be configured with \textit{injectors}, a \textit{scenario}, \textit{validators}, and a \textit{utility function}. For each of them, we defined interfaces that developers have to implement. The simulator orchestrates these artifacts provided by developers to perform the simulation as shown in Figure~\ref{fig:simulation}.

A simulation with the exemplar consists of several rounds while each round corresponds to executing one iteration of the six steps shown in Figure~\ref{fig:simulation}. The number of rounds for a simulation is defined by the \textit{scenario}. The six simulation steps refine the three steps discussed in the introduction and are as follows:

\textit{(1) Inject issues.} 
The simulator injects issues into the \CompArch model and thus, to the architecture of the adaptable software. The \textit{scenario} defines which issues are injected to which elements of the architecture in each simulation round. For each issue type, an \textit{injector} is required that provides the behavior of actually injecting an issue to the element of the architecture defined by the scenario.

\textit{(2) Compute utility}. 
Using the provided \textit{utility function}, the simulator computes the current utility of the adaptable software based on the architecture described by the \CompArch model. Computing the utility after injecting the issues measures the drop in the utility caused by these issues. We use the developing of the utility as one criterion to evaluate adaptation engines.

\textit{(3) Run adaptation engine.}
The adaptation engine developed by the user of the exemplar is executed to perform self-adaptation that aims at resolving the injected issues. For this purpose, the engine analyzes the architecture described by the \CompArch model to identify issues, plans an adaptation of the architecture to resolve these issues, and finally executes the planned adaptation by adjusting the architecture described by the \CompArch model.

\textit{(4) Measure execution time.}
The simulator measures the execution time of the adaptation engine when performing self-adaptation (see previous step). We use the runtime efficiency of a self-adaptation as one criterion to evaluate adaptation engines.

\textit{(5) Validate and update model.} 
The simulator validates the model to check whether issues are remaining in the architecture. As the simulator emulates the adaptable software, it may additionally update the model as a reaction to a valid or invalid self-adaptation (\eg, an adapted architecture might result in changing performance characteristics that should be reflected in the model). Such checks and updates are realized by \textit{validators} that developers provide. 
Moreover, the simulator itself provides generic validators that check architectural constraints (\eg, are all required interfaces of components connected to other components?). 
While the results of such checks give feedback to the developer about the effectiveness of the self-adaptation solution she implemented, the updates of the model enable a simulation over several rounds of self-adaptation.

\textit{(6) Compute utility.}
Similarly to step (2), the utility of the adaptable software is computed, but this time after the self-adaptation (3rd\,step) and when its effects are reflected in the model (5th\,step). Thus, a potential increase of the utility achieved by self-adaptation is measured and used for evaluating the effectiveness of self-adaptation.

All steps except of (3) are performed by the simulator. In each simulation round, the simulator further logs information about the simulation such as the injected issues, computed utility, measured execution time, and the results of the validators. At the very end of the simulation, the simulator stores the raw data of the developing of the utility and execution time values to files and creates corresponding charts. Developers can use the raw data to evaluate in-depth the results and the charts to visually evaluate the effectiveness and efficiency of their self-adaptation solutions.

\subsection{The \CompArch Modeling Language}
\label{subsec:comparch}

To express the architectural runtime model of an adaptable software, we developed a generic yet simple modeling language called \textit{\CompArch} (short for component architecture). 
It supports modeling an individual or a multi-tenant system that comprises many~individual systems (e.g., a marketplace with individual shops).
In this section, we present the \CompArch metamodel, notation, and editor.

\subsubsection{Metamodel} 
\label{subsec:comparch-metamodel}
The \CompArch metamodel depicted in Figure~\ref{fig:metamodel} defines the concepts of the language. Its elements are structured into five groups (see coloring of the elements):

\begin{figure*}
	\centering
	\includegraphics[width=.935\linewidth]{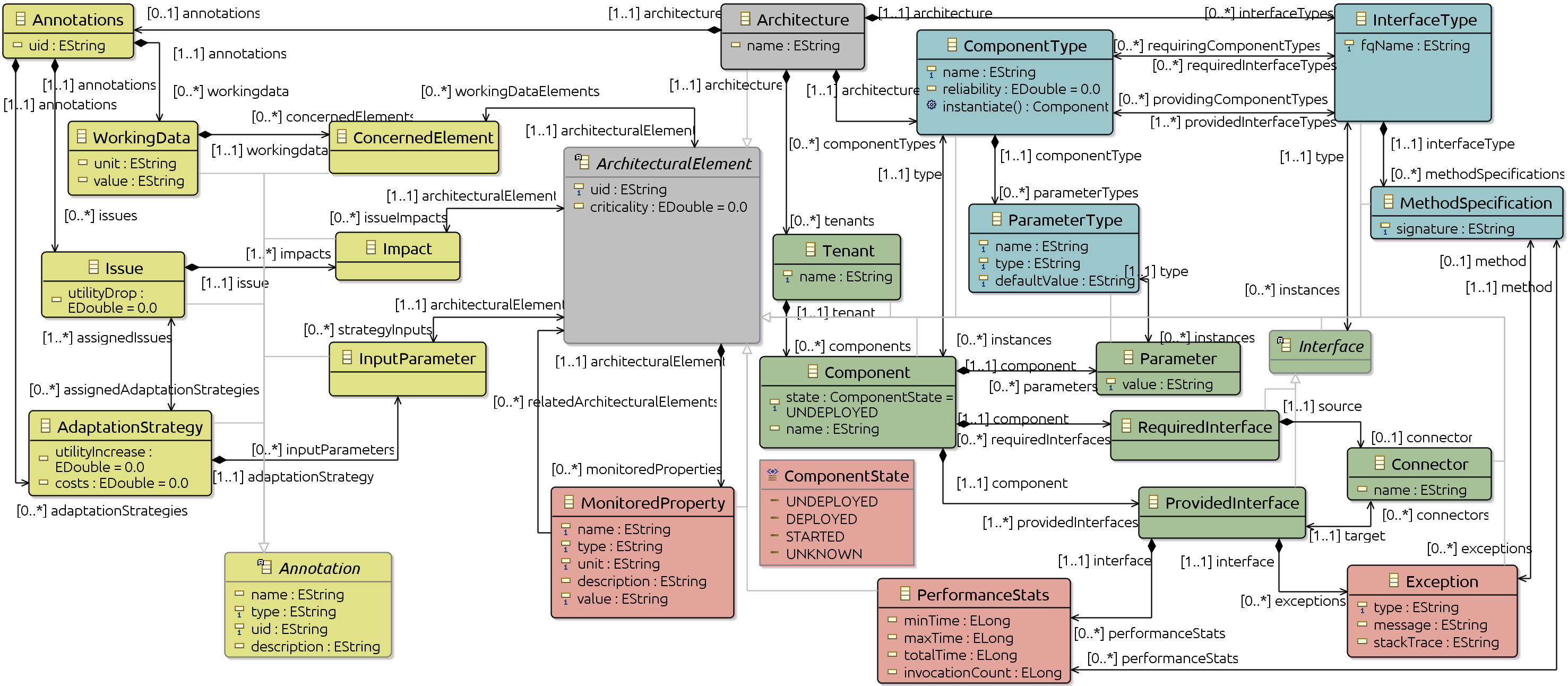}
	\caption{Complete metamodel of the \CompArch modeling language.}
	\label{fig:metamodel}
	\vspace{-1em}
\end{figure*}

\textit{(1) General elements (gray elements).}
The \elem{Architecture} is the root node of the model and represents the modeled system.
The \elem{ArchitecturalElement} is the super class defining an identifier (\elem{uid}) and \elem{criticality} (\ie, how critical the element is for the operation of the system) for all type-, deployment-, and runtime-level elements.

\textit{(2) Type level (blue elements).}
This level covers \elem{ComponentType}s with their required and provided \elem{InterfaceType}s. An \elem{InterfaceType} has a fully qualified name (\elem{fqName}) and contains \elem{MethodSpecification}s, each defined by a \elem{signature}. A \elem{ParameterType} is a configuration option for a \elem{ComponentType} specified by its \elem{name}, primitive data \elem{type}, and \elem{defaultValue}.
A \elem{ComponentType} has a \elem{reliability} and is instantiated to a \elem{Component} by the \elem{instantiate()} operation.

\textit{(3) Deployment level (green elements).}
This level defines the architecture of a deployed system that consists of \elem{Component}s with their \elem{Required-} and \elem{ProvidedInterface}s. Each  component is in a certain life cycle \elem{state} as defined by the enumeration \elem{ComponentState}:
\elem{UNDEPLOYED}, 
\elem{DEPLOYED} (but stopped), 
\elem{STARTED}, and 
\elem{UNKNOWN} (\eg, crashed). 
It is configured by concrete \elem{value}s for its \elem{Parameter}s according to the \elem{ParameterType}s and by \elem{Connector}s that wire their required interfaces to provided interfaces of other components.
Finally, components are associated to \elem{Tenant}s. Each \elem{Tenant} groups the components that comprise the tenant's sub-system in a multi-tenant system. If the modeled system is only an individual system, we consider it as a multi-tenant system with one tenant.

\textit{(4) Runtime level (red elements).}
This level covers runtime concepts that are monitored in the running system.
An \elem{Exception} thrown when using a \elem{ProvidedInterface} is attached to this interface. It is characterized by a \elem{type}, \elem{message}, \elem{stackTrace}, and the \elem{method} that raised the exception.
\elem{PerformanceStats} capture the performance of using a component through a \elem{ProvidedInterface}. For each \elem{method} of the interface, it captures the min., max., and total execution time in \textit{ms} and the number of invocations. Thus, the average execution time is calculated by \elem{totalTime/invocationCount}.~A \elem{MonitoredProperty} describes an application-specific property monitored for an \elem{ArchitecturalElement}. It has a \elem{name}, \elem{description}, data \elem{type}, measurement \elem{unit}, and the monitored \elem{value}.
Finally, the life cycle of \elem{Component}s is monitored and reflected by the \elem{state}~attribute. 

\textit{(5) Annotations (yellow elements).}
\elem{Annotations} to the architecture of the adaptable software are optionally used by adaptation engines to capture knowledge created during self-adaptation (\eg, analysis results and adaptation plans). 
Each \elem{Annotation} is characterized by a \elem{name}, \elem{type}, \elem{uid}, and \elem{description}.  
We consider three categories of annotations: 
\begin{inparaenum}[i)]
\item Metric \elem{WorkingData} with a \elem{value} and measurement \elem{unit} that can refer via \elem{ConcernedElement}s to  \elem{ArchitecturalElement}s 
(\eg, such data may count how often a \elem{Component} violates a performance goal).
\item An \elem{Issue} identified in the architecture and causing a \elem{utilityDrop} of the adaptable software that has \elem{Impact}s on the affected \elem{ArchitecturalElement}s
(\eg, an issue is the violation of a performance goal by a \elem{Component}).
\item An \elem{AdaptationStrategy} with its anticipated \elem{utilityIncrease} and execution \elem{costs} that has been planned to resolve an assigned \elem{Issue} and that may have \elem{ArchitecturalElement}s as \elem{InputParameter}s
(\eg, an adaptation strategy is to replace the \elem{Component} that violates the performance goal).
\end{inparaenum}

\subsubsection{Notation and Modeling Editor}
\label{subsec:editor}

To create and visualize \CompArch models, we developed a notation and a corresponding graphical modeling editor. The notation is based on the proposal by \citet{Kramer&Magee2007} and comprises four types of diagrams.

An \textit{Overview Diagram} lists and supports creating and deleting \elem{ComponentType}s, \elem{InterfaceType}s, and \elem{Tenant}s that are part of the \elem{Architecture}.
An \textit{Interface Type Diagram} supports specifying an \elem{InterfaceType} with its \elem{MethodSpecification}s.
A \textit{Component Type Diagram} supports specifying a \elem{ComponentType} with its \elem{ParameterType}s and associating \elem{InterfaceType}s as required and provided.
A \textit{Tenant Architecture Diagram} supports specifying an architecture for each \elem{Tenant} as shown in Figure~\ref{fig:architecture}, which comprises the \elem{Component}s with their \elem{Required-} and \elem{ProvidedInterface}s, \elem{Parameter}s, and \elem{Connector}s. This diagram furthermore visualizes the runtime-level elements and annotations (\cf red and yellow elements of the metamodel in Figure~\ref{fig:metamodel}) as they refer to architectural elements.

Thus, the editor can be used to create the initial \CompArch~model at design time and to visualize the model at runtime when the model is enriched with runtime information.
The simulator provides snapshots of the \CompArch model on-demand during the simulation.

\subsection{Interacting with \CompArch Models}

Based on \CompArch (\cf~Section~\ref{subsec:comparch}), this section outlines how the simulator and the adaptation engine interact with a \CompArch model during the simulation and self-adaptation (\cf~Figure~\ref{fig:simulator}).

As the simulator emulates the adaptable software, it maintains and updates the model with respect to the deployment- and runtime-level elements (\cf green and red elements in Figure~\ref{fig:metamodel}). Thus, it may update the software architecture (\ie, the structure of \elem{Component}s and \elem{Connector}s, and \elem{Parameter values}) and the runtime information (\ie, \elem{MonitoredProperties}, \elem{PerformanceStats}, and \elem{Exception}s). 
Such updates are performed by injectors when issues are injected in the architecture (\cf first step in Figure~\ref{fig:simulation}). For instance, to emulate a component crash, the \elem{Component} is removed from the architecture, or to emulate a higher load on the software, the \elem{PerformanceStats} values are increased.
Moreover, such updates are performed by validators in response to a self-adaptation (\cf fifth step in Figure~\ref{fig:simulation}). For instance, a self-adaptation reconfigures the architecture of the adaptable software, which changes the performance characteristics and therefore, the simulator updates the \elem{PerformanceStats} values.

Following the relationships defined by the metamodel, an adaptation engine can query and navigate and thus analyze the whole \CompArch model. However, there are restrictions for adapting the model. Focusing on architectural self-adaptation, an adaptation engine can change the composition of \elem{Component}s and \elem{Connector}s by adding, removing, and replacing them (\cf structural adaptation) as well as the operational modes of \elem{Component}s by changing \elem{value}s of configuration \elem{Parameter}s (\cf parameter adaptation). Structural adaptations include changing the life cycle states of \elem{Component}s, for instance, to start or stop a component. Thus, the changeable elements of the model refer to the architecture (\cf green elements in Figure~\ref{fig:metamodel}) while keeping the type level (\cf blue elements) unchanged.
Moreover, the runtime-level elements (\cf red elements) are not changeable as they represent only observable information.
Finally, the annotations (\cf yellow elements) are exclusively used by adaptation engines and can therefore be changed at any will.

\subsection{\CompArch Change Events}
\label{subsec:change-events}

Besides the \CompArch model, the simulator provides on-demand events that notify about changes of the model. We defined eleven types of such events notifying, for instance, about changes of the life cycle state of a \elem{Component}, the addition or removal of a \elem{Component} from the architecture, the re-routing of a \elem{Connector}, the occurrence of an \elem{Exception}, or updates of a \elem{Parameter}, \elem{PerformanceStats}, and \elem{MonitoredProperty}.
Each event has a time stamp of the change and points to the specific architectural element affected by the change.

Thus, adaptation engines can process such events to drive the self-adaptation process (\cf event-condition-action rules for adaptation). Such an event-driven process avoids querying the whole architecture and \CompArch model to identify changes since the events directly point to the changed elements in the architecture. 
This typically results in a runtime-efficient self-adaptation process.

\section{$\text{m}$RUBiS}
\label{sec:mRUBiS}

In this section, we instantiate the generic simulation framework for the self-healing and self-optimization of \mRUBiS by providing injectors, scenarios, validators, and utility functions (\cf~Figure~\ref{fig:framework}).

\subsection{System Description and Goals}
\label{sec:mRUBiS-desc+goals}

\begin{figure}[htb]
	\centering
	\includegraphics[width=.89\linewidth]{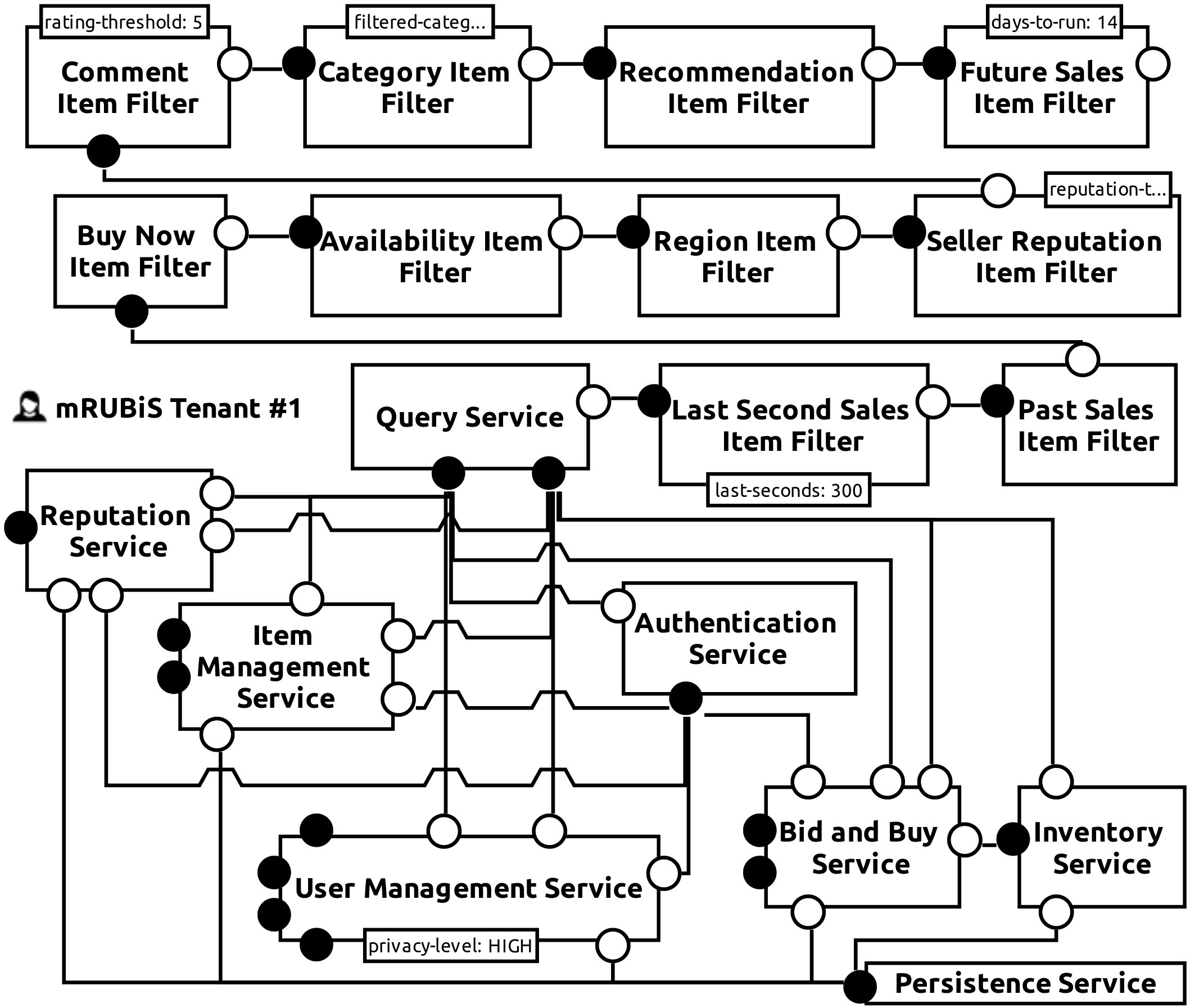}
	\caption{\CompArch model of an individual shop (tenant).}
	\label{fig:architecture}
	\vspace{-1.5em}
\end{figure}

\mRUBiS is a marketplace on which users sell or auction items. It is derived from \textit{RUBiS}, a popular case study to evaluate control theoretic adaptation (\cf~\cite{Patikirikorala+2012}).
Since RUBiS is monolithic and just runs one shop, we added new functionalities (\eg, the pipe of filters), modularized the shop to 18 components, and extended it to a marketplace that hosts arbitrarily many shops. Each of these shops consists of the 18 components and belongs to a tenant (see Figure~\ref{fig:architecture}). All shops share the same component types but each shop has its own, individually configured components. Thus, the architecture of each shop is isolated from the architectures of the other shops (\cf multi-tenancy). While the modularization enables architectural adaptations that are otherwise not possible with the monolithic RUBiS, the multi-tenancy setting enables scaling up the system and thus the architectural runtime model expressed in \CompArch.

As shown in Figure~\ref{fig:architecture}, each shop has components 
to manage items (Item Management Service), users (User Management Service), auctions/purchases (Bid and Buy Service), inventory (Inventory Service), and user ratings (Reputation Service), 
to authenticate users (Authentication Service), and to persist and retrieve data (Persistence and Query Services) from the database.
The pipe of filter components improves the quality of the results when users search for items. The pipe follows the batch sequential pipe-and-filter architectural style. It iteratively filters the list of items obtained from the Query Service by removing items that are not relevant for the specific user and search request. Hence, the pipe improves the quality while also increasing the response time of the search.

The company running \mRUBiS aims for high sales volumes by achieving customer satisfaction and encouraging customers to additional purchases. Therefore, the system should be highly available and its response times should be low. To reach these goals, self-adaptation should be employed to automatically repair failures (\ie, self-healing) and improve performance (\ie, self-optimization).

\subsection{Self-Healing \mRUBiS}
\label{subsec:self-healing}

To achieve a high availability, the self-healing aims at repairing architectural failures that disrupt the operation of \mRUBiS.

\artifact{Injectors}
For the \mRUBiS architecture, we provide injectors for five classes of critical failures (CFs).
They target \elem{Component}s that either crash and enter the \elem{UNKNOWN} life cycle \elem{state} (CF1),
throw \elem{Exception}s exceeding a given threshold (CF2),
are destroyed and removed from the architecture (CF3), and
are continuously affected by CF1 or CF2 requiring novel repair mechanisms (CF4),
and \elem{Connector}s that are lost and removed from the architecture~(CF5).

For these CFs, we consider adaptation strategies (AS) 
that restart (AS1) or redeploy (AS2) the affected \elem{Component} by controlling the component's life cycle \elem{state} (\eg, adjusting the state from \elem{STARTED} to \elem{DEPLOYED} stops the component), 
that replaces the affected \elem{Component} with a new instance of the same (AS3) or different (AS4) \elem{ComponentType}, and 
that reestablishes a lost \elem{Connector}~(AS5).

\artifact{Scenario}
We provide a basic scenario that injects in each simulation round one CF to a random \elem{Component} or \elem{Connector}.

\artifact{Validators}
We provide generic and \mRUBiS-specific validators that check for the validity of the architecture described by the \CompArch model and that take the CFs into account.

\artifact{Utility Function}
We define the utility of \mRUBiS as the sum of the utilities of all \elem{Component}s and the utility of a single component \elem{c} as \elem{c.type.reliability $\times$ c.criticality $\times$ c.connectivity} (\cf~\cite{2017-ICAC}). The connectivity refers to the number of \elem{Connector}s associated to \elem{c}. If a component is affected by a CF, it does not provide any utility so that the utility of \mRUBiS is decreased by the utility of this component.

\artifact{Challenges}
Using this case, developers can investigate different self-adaptation solutions for identifying and resolving CFs.
We provide three example solutions that are either monolithic, decomposed into a MAPE structure, or use the change events (\cf Section~\ref{subsec:change-events}).
Moreover, we used \mRUBiS to scale the size of the architecture and number of injected CFs to investigate the scalability of planning mechanisms~\cite{2017-ICAC}.
In this context, the exemplar provides a generator for \mRUBiS architectures with user-defined numbers of shops to obtain various sizes of the \CompArch model for scalability~analyses.

\subsection{Self-Optimizing \mRUBiS}
\label{subsec:self-optimization}

To achieve low response times, the self-optimization aims at improving the performance of \mRUBiS by architectural reconfiguration.

\artifact{Issues}
We define three performance issues (PIs) with corresponding adaptation strategies (AS) for \mRUBiS that target the pipe of filters (\cf Section~\ref{sec:mRUBiS-desc+goals}). 
The filters of a pipe are ordered based on their performance, that is, filters that perform well are located toward the front of the pipe where more items must be processed than toward the end of the pipe.
In this context, the performance of a filter might change so that the pipe is not optimally ordered (PI1) which requires a re-ordering of filters (AS6).
The average response time of the search in \mRUBiS is above a threshold (PI2) so that filters should be skipped from the pipe (AS7) to reduce the response time at the cost of a decreased quality of the search results.
Similarly, the average response time can be below a threshold (PI3) so that skipped filters can be added back to the pipe (AS8).

\artifact{Scenario}
We provide a basic scenario that injects in each simulation round one PI to the pipe of a random shop of \mRUBiS.

\artifact{Validators}
We provide generic and \mRUBiS-specific validators that check for the validity of the architecture, particularly of the pipe of filters, and that take the PIs into account.

\artifact{Utility Function}
The initial utility of \mRUBiS is computed similarly to the self-healing case. Additionally, it is reduced by 50\% of the utility of a filter that is not located at its optimal position in the pipe, and in total by 20\% if the performance goal is not met (\ie, the average response time of the search is above the threshold).

\artifact{Challenges}
In addition to the challenges for the self-healing case, using this case, developers can investigate self-adaptation solutions that targets a specific architectural style (pipe-and-filter). We provide an example solution that is driven by change events and takes the architectural style into account.
Finally, a challenge is to incorporate the self-healing and self-optimization and thus, multiple adaptation concerns into one self-adaptation solution.

\subsection{Evaluating Self-Adaptation Solutions}
\label{sec:evaluation}

\begin{figure}[t]
	\centering
	\subfloat[Utility of \mRUBiS.\label{fig:utility}]{\includegraphics[width=0.49\linewidth]{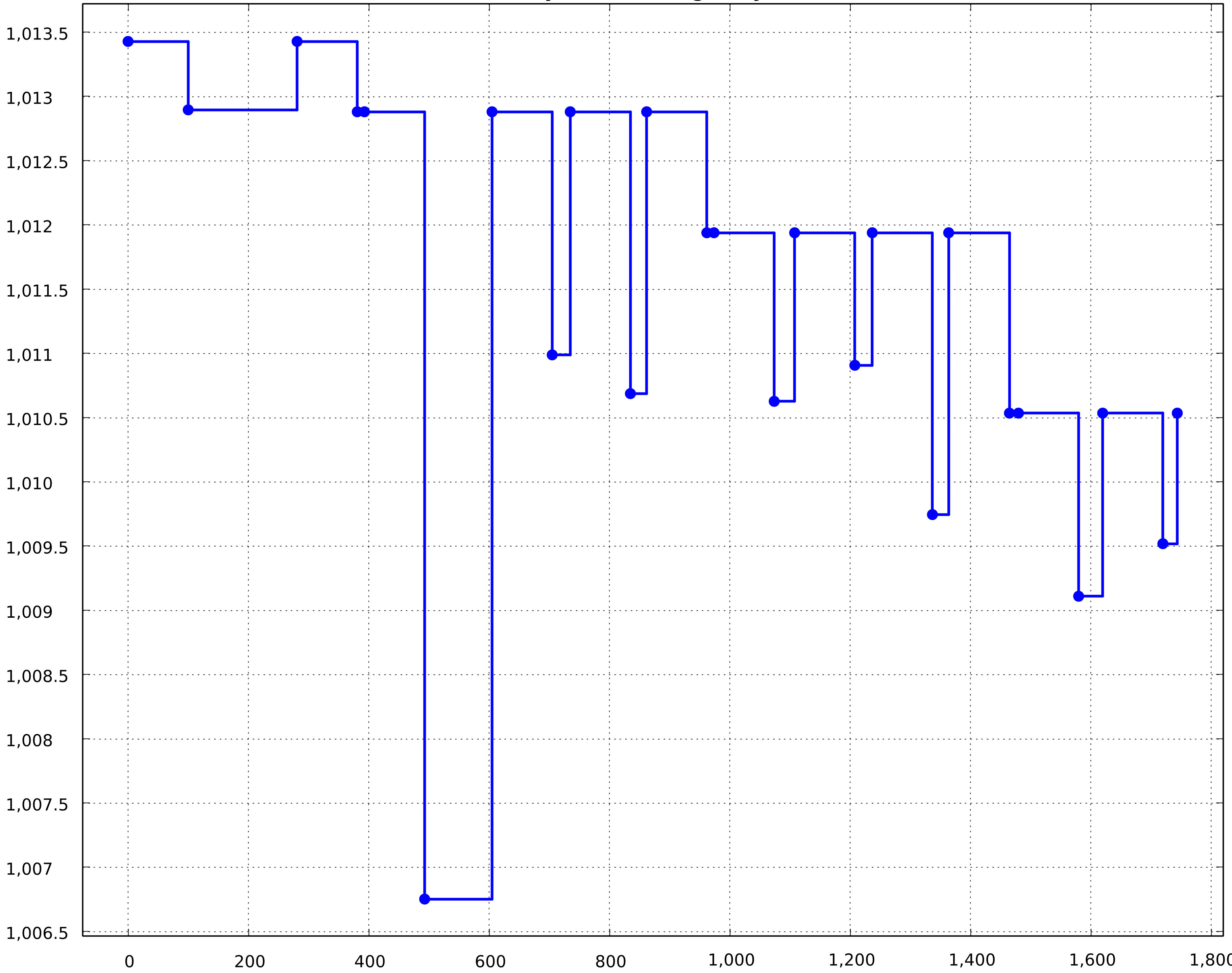}} 
	\hspace{0.01\linewidth}
	\subfloat[Execution time of a solution.\label{fig:executiontime}]{\includegraphics[width=0.46\linewidth]{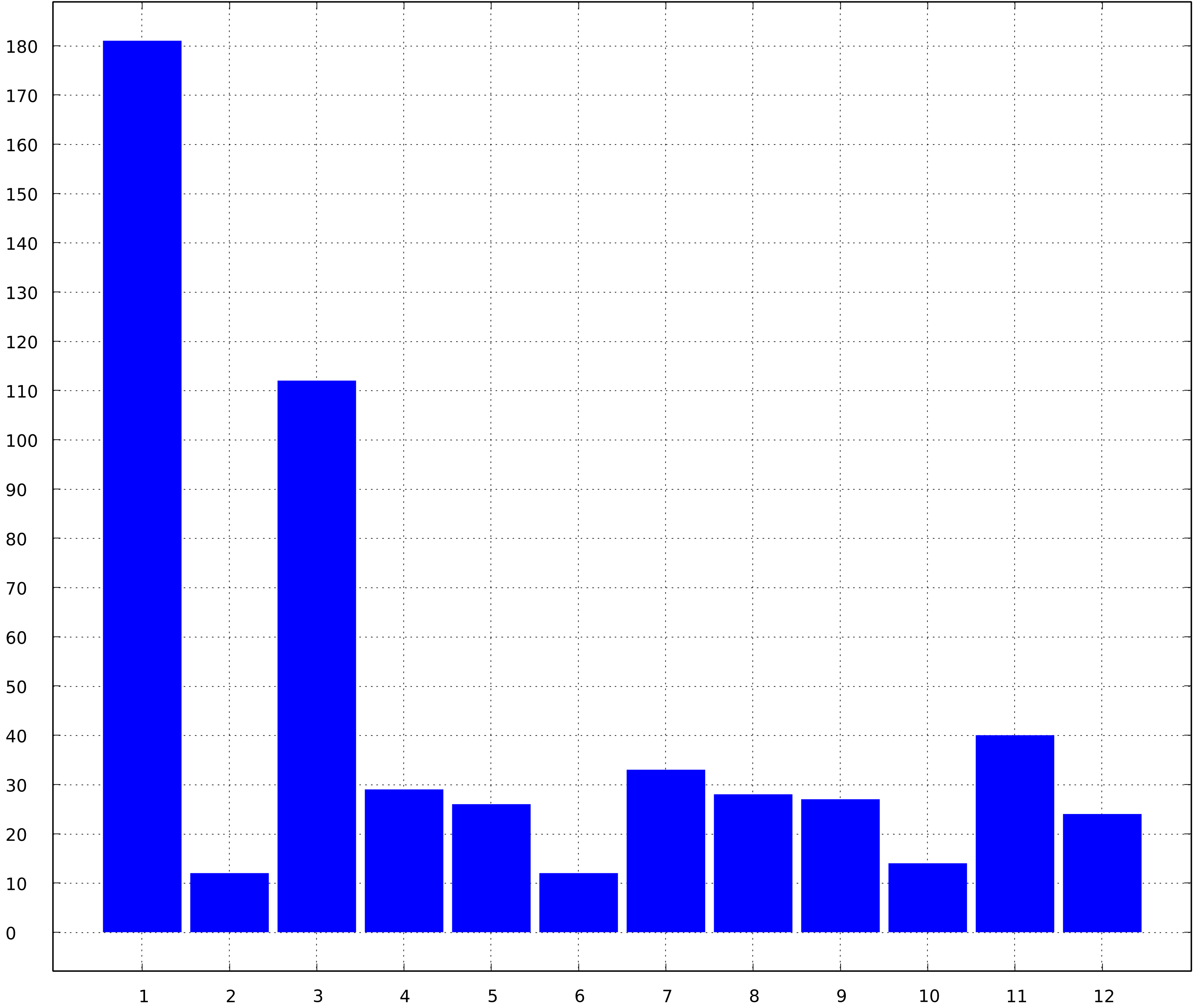}} 
	\caption{Generated charts of the simulation results.}
	\label{fig:charts}
	\vspace{-1em}
\end{figure}

Using the exemplar, self-adaptation solutions are evaluated by the simulation results. The simulator measures the utility of \mRUBiS over time (as calculated by the utility function) and the execution time of a solution in each simulation round.
After the simulation, the simulator generates charts of the results as shown in Figure~\ref{fig:charts} and stores the raw data to csv files.
Developers can use the data and charts to evaluate the effectiveness and efficiency of their solutions.
For instance, Figure~\ref{fig:utility} shows a step chart of the utility over time of \mRUBiS with 50 shops, in which a drop is caused by an injected issue and an increase by a self-adaptation that successfully resolved issues. Since the utility is decreasing in steps, the self-adaptation was not always successful.  
Figure~\ref{fig:executiontime} shows the execution time of a self-adaptation solution (feedback loop) for each simulation round.

Finally, by increasing the number of shops and injected issues per round, the scalability of self-adaptation solutions can be evaluated for large architectures. For this purpose, the exemplar provides a generator for \mRUBiS architectures with arbitrarily many shops.

\section{Implementation}
\label{sec:implementation}

\mRUBiS in terms of the simulator, \CompArch language, and modeling editor is implemented as an Eclipse plug-in on top of the Eclipse Modeling Framework (EMF)~\cite{Steinberg+2008}.
Thus, the exemplar can be used within Eclipse although it can be run decoupled from Eclipse. Using the simulator boils down to using a Java interface to interact with it.
EMF provides the infrastructure for defining the \CompArch language and handling \CompArch models. Thus, users of the exemplar can use Java or any EMF-compatible technology such as OCL and Story Diagrams (\cf~\cite{2017-ICAC,2015-SASOW}) to process a \CompArch model.

\mRUBiS can be easily extended due to its modular implementation architecture. It has sub-projects for 
the generic simulation framework (Section~\ref{subsec:overview}), 
the \CompArch language (Section~\ref{subsec:comparch-metamodel}),
the editor (Section~\ref{subsec:editor}) as well as for the
\mRUBiS-specific extensions to the simulator and for the example solutions (Section~\ref{sec:mRUBiS}).

\section{Conclusion and Future Work}
\label{sec:conclusion}

In this paper we presented \mRUBiS, an extensible exemplar that supports model-based architectural self-adaptation off the shelf. It simulates the adaptable software and provides an architectural runtime model of the software. This model is the interface for adaptation engines to realize and perform architectural self-adaptation. Thus, developers are relieved from implementing a runtime model so that they can focus on developing and evaluating the model-based adaptation logic. 
Moreover, we presented the self-healing and self-optimization case studies for using the exemplar.

So far, \mRUBiS was useful as it supports early testing~\cite{2015-SASOW} and evaluations~\cite{2017-ICAC,2014-TAAS} of solutions. Moreover, we successfully used it for student projects in a course on self-adaptive software, in which the simulator and the students' solutions play against each other using simulation scenarios that were not known to the students.

As future work, we plan 
to improve the efficiency by incrementally computing the utility,
to support environment models,
and to provide further case studies and metrics for evaluating solutions.

\bibliographystyle{ACM-Reference-Format}
\bibliography{SEAMS2018} 
\end{document}